\documentstyle[prl,aps]{revtex}
\begin{document}
\draft
\title{Realization of an $n$-qubit controlled-$U$ gate with superconducting quantum
interference devices or atoms in cavity QED}
\author{Chui-Ping Yang and Siyuan Han}
\address{Department of Physics and Astronomy, University of Kansas, Lawrence,\\
Kansas 66045}
\maketitle

\begin{abstract}
We propose an approach to realize an $n$-qubit controlled-$U$ gate with
superconducting quantum interference devices (SQUIDs) in cavity QED. In this
approach, the two lowest levels of a SQUID represent the two logical states
of a qubit while a higher-energy intermediate level serves the gate
manipulation. Our method operates essentially by creating a single photon
through one of the control SQUIDs and then performing an arbitrary unitrary
transformation on the target SQUID with the assistance of the cavity photon.
In addition, we show that the method can be applied to implement an $n $%
-qubit controlled-$U$ gate with atomic qubits in cavity QED.
\end{abstract}

\date{\today }


\begin{center}
{\bf I. INTRODUCTION }
\end{center}

Superconducting devices such as Cooper pair boxes, Josephson junctions, and
superconducting quantum interference devices (SQUIDs) have attracted much
attention in the quantum-information community. Because they are relatively
easy to scale up and have been demonstrated to have relatively long
decoherence times [1-7], they have been considered as promising candidates
for physical implementation of quantum computing. It is known that the
building blocks of quantum computers are single-qubit logic gates and
two-qubit logic gates [8]. In the past few years, for SQUID systems, many
methods for realizing a single-qubit arbitrary rotation gate and a two-qubit
controlled-NOT (or controlled-phase shift) gate have been presented [9-16].

Multiqubit controlled gates play an important role in constructing quantum
computational networks and realizing quantum error correction protocols and
implementing quantum algorithms. Recently, much attention is paid to
physical realization of multiqubit controlled gates [17-19]. It is known
that when using the conventional gate-decomposition protocols to construct a
multiqubit controlled gate [20-22], the procedure usually becomes
complicated as the number of qubits increases. Therefore, it is important to
find a more efficient way to implement multiqubit controlled gates.

In this paper we focus on how to realize an $n$-qubit controlled-$U$ gate
with $n$ SQUIDs ($1,2,...,n$) based on cavity QED. Recently, it has been
predicted that the strong coupling limit of cavity QED, which is difficult
to achieve with atoms in a microwave cavity, can readily be realized with
superconducting charge qubits [23,24], superconducting flux qubits [25], or
semiconducting quantum dots [26]. And more recently, the strong coupling
cavity QED has been experimentally demonstrated with superconducting charge
qubits and flux qubits [27,28] and semiconductor quantum dots embedded in a
microcavity [29-31]. The controlled gate considered in this paper is shown
in Fig. 1, which is defined as follows: (a) It contains $n-1$ control qubits
and one target qubit, (b) It leaves the state of the target qubit unchanged
if not all the control qubits are in the state $\left| 1\right\rangle $. (c)
However, when all control qubits are in the state $\left| 1\right\rangle ,$
an arbitrary unitary transformation $U$ is performed on the target qubit.

To implement the general multiqubit controlled gate described above, three
levels $\left| 0\right\rangle ,$ $\left| 1\right\rangle ,$ and $\left|
2\right\rangle $ of each SQUID will be employed (Fig. 2). For each SQUID,
the two lowest levels $\left| 0\right\rangle $ and $\left| 1\right\rangle $
represent two logical states of a qubit while the higher-energy level $%
\left| 2\right\rangle $ is used to facilitate coherent control and
manipulation of quantum states of the qubit. The method presented here
operates essentially by (a) creating a single photon through one of the
control SQUIDs, (b) performing a general $U$ on the target SQUID with the
aid of the cavity photon as follows [32] 
\begin{equation}
U=e^{i\alpha }R_z\left( \beta \right) R_y\left( \gamma \right) R_z\left(
\delta \right) ,
\end{equation}
and (c) finally performing operations to have the cavity mode return to its
original vacuum state. In equation (1), $\alpha ,\beta ,\gamma ,$ and $%
\delta $ are arbitrary real numbers, $R_y$ and $R_z$ represent rotations
along $y$ and $z$ axes on a Bloch sphere, which are described by matrices 
\begin{eqnarray}
R_y\left( \gamma \right) &=&\left( 
\begin{array}{cc}
\cos \frac \gamma 2 & -\sin \frac \gamma 2 \\ 
\sin \frac \gamma 2 & \cos \frac \gamma 2
\end{array}
\right) , \\
R_z\left( \vartheta \right) &=&\left( 
\begin{array}{cc}
e^{-i\vartheta /2} & 0 \\ 
0 & e^{i\vartheta /2}
\end{array}
\right) ,\vartheta =\beta ,\delta ,
\end{eqnarray}
in a single-qubit computational subspace formed by the two logical states $%
\left| 0\right\rangle =\left( 1,0\right) ^T$ and $\left| 1\right\rangle
=\left( 0,1\right) ^T$ of the target qubit.

As shown below, this scheme has the following advantages: (i) No auxiliary
SQUIDs or measurement is needed during the entire operation, thus the
hardware resources is reduced and the operation is simplified; (ii) As
tunneling between the qubit levels $\left| 0\right\rangle $ and $\left|
1\right\rangle $ is not required during the operation, decay from the level $%
\left| 1\right\rangle $ can be made negligibly small during the operation
(via prior adjustment of the potential barrier between the qubit levels $%
\left| 0\right\rangle $ and $\left| 1\right\rangle $ [33]) and therefore the
storage time of each qubit can be made much longer; (iii) The coupling
constants of SQUIDs with the cavity mode could be different, hence neither
identical SQUIDs nor exact placement of SQUIDs is needed; and (v) More
importantly, the gate operations are significantly simplified as the number
of qubits increases, when compared with the conventional gate-decomposition
protocols. In addition, it is interesting to note that the method can
readily be extended to obtain an $n$-qubit controlled-$U$ with atomic qubits
in cavity QED.

This paper is outlined as follows. In Sec. II, we review the basic theory of
a SQUID coupled to a single-mode cavity or driven by a classical microwave
pulse. In Sec. III, we show a way to realize a two-qubit controlled-$U$ gate
with two SQUIDs coupled to a cavity. In Sec. IV, we discuss how to extend
the method to achieve an $n$-qubit controlled-$U$ gate with $n$ SQUIDs in
cavity QED. In Sec. V, we compare the present method with conventional
gate-construction protocols. In Sec. VI, we give a brief discussion on
experimental issues for the realization of an $n$-qubit controlled-rotation
gate. In Sec. VII, we further show how to apply our method to implementing
an $n$-qubit controlled-$U$ gate with $n$ atoms using one cavity only. A
concluding summary is given in the last section.

\begin{center}
{\bf II. BASIC THEORY}
\end{center}

The SQUIDs throughout this paper are rf SQUIDs each consisting of a
Josephson tunnel junction in a superconducting loop (typical size of an rf
SQUID is on the order of 10 $-$ 100 $\mu $m). The Hamiltonian of an rf SQUID
(with junction capacitance $C$ and loop inductance $L$) has the usual form
[5] 
\begin{equation}
H_s=\frac{Q^2}{2C}+\frac{\left( \Phi -\Phi _x\right) ^2}{2L}-E_J\cos \left(
2\pi \frac \Phi {\Phi _0}\right) ,
\end{equation}
where $\Phi $ is the magnetic flux threading the ring, $Q$ is the total
charge on the capacitor, $\Phi _x$ is the external magnetic flux applied to
the ring, and $E_J$ $\equiv I_c\Phi _0/2\pi $ is the maximum Josephson
coupling energy ($I_c$ is the critical current of the junction and $\Phi
_0=h/2e$ is the flux quantum).

\begin{center}
{\bf A. SQUID-cavity resonant interaction}
\end{center}

Consider a SQUID coupled to a single-mode microwave cavity field. The SQUID
is biased properly to have a $\Lambda $-type configuration formed by three
lowest levels, denoted by $\left| 0\right\rangle ,$ $\left| 1\right\rangle $
and $\left| 2\right\rangle $ with energy eigenvalues $E_0,$ $E_1,$ and $E_2$%
, respectively (Fig. 2). The transition frequency between the two levels $%
\left| i\right\rangle $ and $\left| j\right\rangle $ is $\nu _{ij}\equiv
\omega _{ij}/\left( 2\pi \right) =\left| E_i-E_j\right| /\hbar $ $(i,j\in
\left\{ 0,1,2\right\} ,i\neq j).$ Suppose that the coupling of $\left|
0\right\rangle ,\left| 1\right\rangle $ and $\left| 2\right\rangle $ with
other levels of the SQUID via the cavity is negligible, which can readily be
achieved by adjusting the level spacings of the SQUID [33]. We can show that
when the cavity mode is resonant with the $\left| 0\right\rangle
\leftrightarrow \left| 2\right\rangle $ transition while decoupled (highly
detuned) from the $\left| 1\right\rangle \leftrightarrow \left|
2\right\rangle $ transition and the $\left| 0\right\rangle \leftrightarrow
\left| 1\right\rangle $ transition of the SQUID, the interaction Hamiltonian
in the interaction picture, after the rotating-wave approximation, is
described by [12] 
\begin{equation}
H_I=\hbar \left( ga^{+}\left| 0\right\rangle \left\langle 2\right| +\text{%
h.c.}\right) ,
\end{equation}
where $a^{+}$ and $a$ are the creation and annihilation operators of the
cavity mode, and $g$ is the coupling constant between the cavity mode and
the $\left| 0\right\rangle \leftrightarrow \left| 2\right\rangle $
transition of the SQUID. For a superconducting one-dimensional transmission
line standing-wave cavity, $g$ is given by 
\begin{equation}
\hbar g\left( x\right) =\frac{M_{sc}}L\sqrt{\frac{h\nu _c}{L_0l}}%
\left\langle 0\right| \Phi \left| 2\right\rangle \sin \left( \frac{2\pi }%
\lambda x\right) ,
\end{equation}
where $M_{sc}$ is the SQUID-cavity mutual inductance, $L_0$ is the
inductance per unit length of the cavity, $l$ is the length of the cavity, $%
\nu _c\equiv \omega _c/\left( 2\pi \right) $ is the frequency of the cavity
mode with wavelength $\lambda ,$ and $x$ is position of the center of the
SQUID in the cavity.

In the case when the cavity is initially in the photon-number state $\left|
n\right\rangle $, the time evolution of the states of the system, under the
Hamiltonian (5), is as follows 
\begin{eqnarray}
\left| 0\right\rangle \left| n\right\rangle &\rightarrow &\cos \sqrt{n}%
gt\left| 0\right\rangle \left| n\right\rangle -i\sin \sqrt{n}gt\left|
2\right\rangle \left| n-1\right\rangle ,  \nonumber \\
\left| 2\right\rangle \left| n\right\rangle &\rightarrow &-i\sin \sqrt{n+1}%
gt\left| 0\right\rangle \left| n+1\right\rangle +\cos \sqrt{n+1}gt\left|
2\right\rangle \left| n\right\rangle .
\end{eqnarray}
SQUIDs may have non-uniform device parameters and/or be not exactly placed
in the cavity. Therefore, the coupling strength $g$ may not be identical for
different SQUIDs. In the following, we will replace $g$ by $g_1,$ $g_2,...,$
and $g_n$ for SQUIDs $1,2,...,$ and $n,$ respectively.{\bf \ }

As shown below, resonant interaction between the cavity mode and the $\left|
1\right\rangle \leftrightarrow \left| 2\right\rangle $ transition of the
SQUID is needed. In this case, the time evolution for the states of the
system is similar to Eq. (7). One just needs to replace $\left|
0\right\rangle $ in Eq. (7) by $\left| 1\right\rangle .$ The coupling
constant of the cavity mode with the $\left| 1\right\rangle \leftrightarrow
\left| 2\right\rangle $ transition of the SQUID can be set to be the same as
that with the $\left| 0\right\rangle \leftrightarrow \left| 2\right\rangle $
transition of the same SQUID by simply changing the external flux bias from $%
\left( 0.5-\delta \right) \Phi _0$ to $\left( 0.5+\delta \right) \Phi _0$
(see Fig. 3).

\begin{center}
{\bf B. SQUID-cavity off resonant interaction}
\end{center}

Consider a system composed of a SQUID and a single-mode cavity. Suppose that
the cavity mode is off resonant with the $\left| 0\right\rangle
\leftrightarrow \left| 2\right\rangle $ transition (i.e., $\Delta =\omega
_{20}-\omega _c>>\widehat{g}$) while decoupled from the $\left|
1\right\rangle \leftrightarrow \left| 2\right\rangle $ transition and the $%
\left| 0\right\rangle \leftrightarrow \left| 1\right\rangle $ transition of
the SQUID [Fig. 4(a)]. Here, $\Delta $ is the detuning between the
cavity-mode frequency and the $\left| 0\right\rangle \leftrightarrow \left|
2\right\rangle $ transition frequency of the SQUID, and $\widehat{g}$ is the
coupling constant of the cavity mode with the $\left| 0\right\rangle
\leftrightarrow \left| 2\right\rangle $ transition of the SQUID. Under this
condition, the SQUID has a negligible probability of making a transition
between the ground level $\left| 0\right\rangle $ and the excited level $%
\left| 2\right\rangle .$ Therefore, the effective interaction Hamiltonian in
the interaction picture can be written as [34] 
\begin{equation}
H_e=\hbar \frac{\widehat{g}^2}\Delta \left( \left| 2\right\rangle
\left\langle 2\right| -\left| 0\right\rangle \left\langle 0\right| \right)
a^{+}a.
\end{equation}

From the Hamiltonian (8), it is straightforward to see that if the cavity
mode is initially in the photon-number state $\left| n\right\rangle $, the
time evolution of the states of the system is then given by 
\begin{eqnarray}
\left| 0\right\rangle \left| n\right\rangle &\rightarrow &e^{i\frac{\widehat{%
g}^2}\Delta nt}\left| 0\right\rangle \left| n\right\rangle ,  \nonumber \\
\left| 2\right\rangle \left| n\right\rangle &\rightarrow &e^{-i\frac{%
\widehat{g}^2}\Delta nt}\left| 2\right\rangle \left| n\right\rangle .
\end{eqnarray}

In the following, we will need to use off-resonant interaction between the
cavity mode and the $\left| 1\right\rangle \leftrightarrow \left|
2\right\rangle $ transition of the SQUID [Fig. 4(b)]. In this case, the time
evolution for the states of the system is similar to Eq. (9). One just needs
to replace $\left| 0\right\rangle $ in Eq. (9) by $\left| 1\right\rangle .$
The detuning is given by $\Delta =\omega _{21}-\omega _c.$ $\widehat{g}$ is
the coupling constant between the cavity mode and the $\left| 1\right\rangle
\leftrightarrow \left| 2\right\rangle $ transition of the SQUID. As
described above, i.e., by changing the external flux bias from $\left(
0.5-\delta \right) \Phi _0$ to $\left( 0.5+\delta \right) \Phi _0,$ both $%
\Delta $ and $\widehat{g}$ can be set to be the same as those for the case
of the cavity mode being off-resonant with the $\left| 0\right\rangle
\leftrightarrow \left| 2\right\rangle $ transition of the same SQUID (Fig.
4).

\begin{center}
{\bf C. SQUID-microwave resonant interaction}
\end{center}

In this section, we consider a SQUID driven by a classical microwave pulse
with the magnetic component ${\bf B}_{\mu w}({\bf r,t})={\bf B}_{\mu w}({\bf %
r})$ cos $\left( \omega _{\mu w}t+\phi \right) $. Here, ${\bf B}_{\mu w}(%
{\bf r),}$ $\omega _{\mu w}$, and $\phi $ are the magnetic field amplitude,
carrier frequency, and phase of the microwave pulse. Assume that the
microwave pulse is resonant with the $\left| 1\right\rangle \leftrightarrow
\left| 2\right\rangle $ transition but decoupled from the $\left|
0\right\rangle \leftrightarrow \left| 2\right\rangle $ transition and the $%
\left| 0\right\rangle \leftrightarrow \left| 1\right\rangle $ transition of
the SQUID. The interaction Hamiltonian in the interaction picture is then
given by 
\begin{equation}
H_I=\frac \hbar 2\left( \Omega e^{i\phi }\left| 1\right\rangle \left\langle
2\right| +\text{h.c.}\right) ,
\end{equation}
where $\Omega $ is the Rabi frequency of the pulse, which has the following
form [12] 
\begin{equation}
\Omega \left( t\right) =\frac 1{L\hbar }\left\langle 1\right| \Phi \left|
2\right\rangle \int_S{\bf B}_{\mu w}({\bf r})\cdot d{\bf S.}
\end{equation}
From the Hamiltonian (10), it is easy to find the following state rotation 
\begin{eqnarray}
\left| 1\right\rangle &\rightarrow &\cos \frac \Omega 2t\left|
1\right\rangle -ie^{-i\phi }\sin \frac \Omega 2t\left| 2\right\rangle , 
\nonumber \\
\left| 2\right\rangle &\rightarrow &-ie^{i\phi }\sin \frac \Omega 2t\left|
1\right\rangle +\cos \frac \Omega 2t\left| 2\right\rangle .
\end{eqnarray}

\begin{center}
{\bf III. TWO-QUBIT CONTROLLED-}$U${\bf \ GATES}
\end{center}

For two qubits, there are four computational basis states denoted by $\left|
00\right\rangle ,\left| 01\right\rangle ,\left| 10\right\rangle ,$ and $%
\left| 11\right\rangle ,$ respectively. A two-qubit controlled-$U$ gate is
described by 
\begin{eqnarray}
\left| 00\right\rangle &\rightarrow &\left| 00\right\rangle ,  \nonumber \\
\;\left| 01\right\rangle &\rightarrow &\left| 01\right\rangle ,  \nonumber \\
\left| 10\right\rangle &\rightarrow &\left| 1\right\rangle U\left|
0\right\rangle ,  \nonumber \\
\;\left| 11\right\rangle &\rightarrow &\left| 1\right\rangle U\left|
1\right\rangle ,
\end{eqnarray}
which implies that if and only if the control qubit (the first qubit) is in
the state $\left| 1\right\rangle $, a unitary transformation $U$ is
performed on the target qubit (the second qubit) and nothing happens
otherwise.

Now let us discuss how to obtain the two-qubit controlled-$U$ gate (13) with
two SQUIDs $1$ and $2$ coupled to a microwave cavity. The SQUIDs considered
here have the $\Lambda $-type three-level configuration as depicted in Fig.
2. The transition between any two levels for each SQUID is initially
decoupled from the cavity mode (e.g., via prior adjustment of the level
spacings). And the cavity mode is initially in the vacuum state $\left|
0\right\rangle _c.$ To realize the gate (13), it is required to perform the
following transformation 
\begin{eqnarray}
\left| 0\right\rangle _1\left| 0\right\rangle _2\otimes \left|
0\right\rangle _c &\rightarrow &\left| 0\right\rangle _1\left|
0\right\rangle _2\otimes \left| 0\right\rangle _c,  \nonumber \\
\;\left| 0\right\rangle _1\left| 1\right\rangle _2\otimes \left|
0\right\rangle _c &\rightarrow &\left| 0\right\rangle _1\left|
1\right\rangle _2\otimes \left| 0\right\rangle _c,  \nonumber \\
\left| 1\right\rangle _1\left| 0\right\rangle _2\otimes \left|
0\right\rangle _c &\rightarrow &\left| 1\right\rangle _1\left[ e^{i\alpha
}R_z\left( \beta \right) R_y\left( \gamma \right) R_z\left( \delta \right)
\left| 0\right\rangle _2\right] \otimes \left| 0\right\rangle _c,  \nonumber
\\
\;\left| 1\right\rangle _1\left| 1\right\rangle _2\otimes \left|
0\right\rangle _c &\rightarrow &\left| 1\right\rangle _1\left[ e^{i\alpha
}R_z\left( \beta \right) R_y\left( \gamma \right) R_z\left( \delta \right)
\left| 1\right\rangle _2\right] \otimes \left| 0\right\rangle _c,
\end{eqnarray}
where subscripts $1$ and $2$ represent SQUID $1$ (the control qubit) and
SQUID $2$ (the target qubit), respectively.

We note that the unitary transformations, involved in the last two lines of
Eq. (14), can be realized through the following operation sequence:

First, creating a single photon through SQUID $1$ as follows 
\begin{equation}
\left| 1\right\rangle _1\left| 0\right\rangle _c\rightarrow \left|
0\right\rangle _1\left| 1\right\rangle _c.
\end{equation}

Second, performing rotations $R_z(\delta ),$ $R_y(\gamma ),$ and then $%
R_z(\beta )$ on the states of SQUID $2$ with the assistance of the photon,
i.e., 
\begin{equation}
R_z(\delta ): 
\begin{array}{c}
\left| 0\right\rangle _2\left| 1\right\rangle _c\rightarrow e^{-i\delta
}\left| 0\right\rangle _2\left| 1\right\rangle _c, \\ 
\left| 1\right\rangle _2\left| 1\right\rangle _c\rightarrow e^{i\delta
}\left| 1\right\rangle _2\left| 1\right\rangle _c,
\end{array}
\end{equation}
\begin{equation}
R_y(\gamma ): 
\begin{array}{c}
\left| 0\right\rangle _2\left| 1\right\rangle _c\rightarrow \left( \cos 
\frac \gamma 2\left| 0\right\rangle _2+\sin \frac \gamma 2\left|
1\right\rangle _2\right) \left| 1\right\rangle _c, \\ 
\left| 1\right\rangle _2\left| 1\right\rangle _c\rightarrow \left( -\sin 
\frac \gamma 2\left| 0\right\rangle _2+\cos \frac \gamma 2\left|
1\right\rangle _2\right) \left| 1\right\rangle _c,
\end{array}
\end{equation}
\begin{equation}
R_z(\beta ): 
\begin{array}{c}
\left| 0\right\rangle _2\left| 1\right\rangle _c\rightarrow e^{-i\beta
}\left| 0\right\rangle _2\left| 1\right\rangle _c, \\ 
\left| 1\right\rangle _2\left| 1\right\rangle _c\rightarrow e^{i\beta
}\left| 1\right\rangle _2\left| 1\right\rangle _c.
\end{array}
\end{equation}

Third, performing a phase-shift $e^{i\alpha }$ on the states of SQUID $2$
with the aid of the photon, i.e., 
\begin{eqnarray}
\left| 0\right\rangle _2\left| 1\right\rangle _c &\rightarrow &e^{i\alpha
}\left| 0\right\rangle _2\left| 1\right\rangle _c, \\
\left| 1\right\rangle _2\left| 1\right\rangle _c &\rightarrow &e^{i\alpha
}\left| 1\right\rangle _2\left| 1\right\rangle _c.  \nonumber
\end{eqnarray}

Last, returning SQUID $1$ and the cavity mode to their original states,
i.e., 
\begin{equation}
\left| 0\right\rangle _1\left| 1\right\rangle _c\rightarrow \left|
1\right\rangle _1\left| 0\right\rangle _c.
\end{equation}

In the following, we will list operations required for the realization of
the above transformations (15)-(20). {\bf \ }

Step (i): Apply a $\pi $ microwave pulse ($\Omega \tau _{\mu w}=\pi ,$ where 
$\tau _{\mu w}$ is the pulse duration) with $\phi =-\pi /2$ to SQUID $1$
[Fig. 5(a)]. The pulse is resonant with the $\left| 1\right\rangle
\leftrightarrow \left| 2\right\rangle $ transition of SQUID $1.$ After the
pulse, the transformation $\left| 1\right\rangle \rightarrow \left|
2\right\rangle $ of SQUID $1$ is obtained.

Step (ii): Bring the $\left| 0\right\rangle \leftrightarrow \left|
2\right\rangle $ transition of SQUID $1$ to resonance with the cavity mode
for an interaction time $\tau _1=\pi /\left( 2g_1\right) $ [Fig. 5(b)],
resulting in $\left| 2\right\rangle _1\left| 0\right\rangle _c\rightarrow
-i\left| 0\right\rangle _1\left| 1\right\rangle _c$.

After the operations of Step (i) and Step (ii), the transformation (15) is
obtained as follows

\begin{equation}
\left| 1\right\rangle _1\left| 0\right\rangle _c\stackrel{\text{(i)}}{%
\rightarrow }\left| 2\right\rangle _1\left| 0\right\rangle _c\stackrel{\text{%
(ii)}}{\rightarrow }-i\left| 0\right\rangle _1\left| 1\right\rangle _c
\end{equation}
up to a phase factor $-i,$ which is inevitable according to Eq. (7) but can
be removed by introducing a phase factor $i$ to the transformation (20) (see
below).

Step (iii): Adjust the level structure of SQUID $2$ and apply a $\pi $
microwave pulse with $\phi =-\pi /2$ to SQUID $2$ [Fig. 5(c)]. The pulse is
resonant with the $\left| 1\right\rangle \leftrightarrow \left|
2\right\rangle $ transition of SQUID $1,$ leading to the transformation $%
\left| 1\right\rangle \rightarrow \left| 2\right\rangle $ of SQUID $2.$

Step (iv): Adjust the level structure of SQUID $2$ to obtain off-resonant
interaction between the cavity mode and the $\left| 0\right\rangle
\leftrightarrow \left| 2\right\rangle $ transition of SQUID $2$ [Fig. 5(d)].
After an interaction time $\tau _2=\delta \widetilde{\Delta }/\widetilde{g}%
^2,$ the state $\left| 0\right\rangle _2\left| 1\right\rangle _c$ goes to $%
e^{-i\delta }\left| 0\right\rangle _2\left| 1\right\rangle _c$ while the
state $\left| 2\right\rangle _2\left| 1\right\rangle _c$ changes to $%
e^{i\delta }\left| 2\right\rangle _2\left| 1\right\rangle _c.$

Step (v): Repeat the operation of Step (iii) but set $\phi =\pi /2,$ leading
to the transformation $\left| 2\right\rangle \rightarrow \left|
1\right\rangle $ of SQUID $2.$

It is easy to see that after the operations of Steps (iii)-(v), the
transformation (16) is implemented as follows: 
\begin{eqnarray}
&&\left| 0\right\rangle _2\left| 1\right\rangle _c\stackrel{\text{(iii)}}{%
\rightarrow }\left| 0\right\rangle _2\left| 1\right\rangle _c\stackrel{\text{%
(iv)}}{\rightarrow }e^{-i\delta }\left| 0\right\rangle _2\left|
1\right\rangle _c\stackrel{\text{(v)}}{\rightarrow }e^{-i\delta }\left|
0\right\rangle _2\left| 1\right\rangle _c,  \nonumber \\
&&\left| 1\right\rangle _2\left| 1\right\rangle _c\stackrel{\text{(iii)}}{%
\rightarrow }\left| 2\right\rangle _2\left| 1\right\rangle _c\stackrel{\text{%
(iv)}}{\rightarrow }e^{i\delta }\left| 2\right\rangle _2\left|
1\right\rangle _c\stackrel{\text{(v)}}{\rightarrow }e^{i\delta }\left|
1\right\rangle _2\left| 1\right\rangle _c,
\end{eqnarray}

Step (vi) Bring the $\left| 1\right\rangle \leftrightarrow \left|
2\right\rangle $ transition of SQUID $2$ to resonance with the cavity mode
for an interaction time $\tau _3=\pi /\left( 2g_2\right) $ [Fig. 5(e)]. As a
result, the state $\left| 0\right\rangle _2\left| 1\right\rangle _c$ remains
unchanged while the state $\left| 1\right\rangle _2\left| 1\right\rangle _c$
changes to $-i\left| 2\right\rangle _2\left| 0\right\rangle _c.$

Step (vii): Bring the $\left| 0\right\rangle \leftrightarrow \left|
2\right\rangle $ transition of SQUID $2$ to resonance with the cavity mode
for an interaction time $\tau _4=\gamma /\left( 2g_2\right) $ [Fig. 5(f)],
resulting in $\left| 0\right\rangle _2\left| 1\right\rangle _c\rightarrow
\cos \frac \gamma 2\left| 0\right\rangle _2\left| 1\right\rangle _c-i\sin 
\frac \gamma 2\left| 2\right\rangle _2\left| 0\right\rangle _c$ and $\left|
2\right\rangle _2\left| 0\right\rangle _c\rightarrow -i\sin \frac \gamma 2%
\left| 0\right\rangle _2\left| 1\right\rangle _c+\cos \frac \gamma 2\left|
2\right\rangle _2\left| 0\right\rangle _c.$

Step (viii): Bring the $\left| 1\right\rangle \leftrightarrow \left|
2\right\rangle $ transition of SQUID $2$ to resonance with the cavity mode
for an interaction time $\tau _5=3\pi /\left( 2g_2\right) $ [Fig. 5(e)]. As
a result, the states $\left| 0\right\rangle _2\left| 1\right\rangle _c$
remains unchanged while the state $\left| 2\right\rangle _2\left|
0\right\rangle _c$ becomes $i\left| 1\right\rangle _2\left| 1\right\rangle
_c.$

One can see that after the operations of Steps (vi)-(viii), the
transformation (17) is obtained as follows:

\begin{eqnarray*}
&&\left| 0\right\rangle _2\left| 1\right\rangle _c\stackrel{\text{(vi)}}{%
\rightarrow }\left| 0\right\rangle _2\left| 1\right\rangle _c\stackrel{\text{%
(vii)}}{\rightarrow }\cos \frac \gamma 2\left| 0\right\rangle _2\left|
1\right\rangle _c-i\sin \frac \gamma 2\left| 2\right\rangle _2\left|
0\right\rangle _c\stackrel{\text{(viii)}}{\rightarrow }\left( \cos \frac 
\gamma 2\left| 0\right\rangle _2+\sin \frac \gamma 2\left| 1\right\rangle
_2\right) \left| 1\right\rangle _c, \\
&&\left| 1\right\rangle _2\left| 1\right\rangle _c\stackrel{\text{(vi)}}{%
\rightarrow }-i\left| 2\right\rangle _2\left| 0\right\rangle _c\stackrel{%
\text{(vii)}}{\rightarrow }-\sin \frac \gamma 2\left| 0\right\rangle
_2\left| 1\right\rangle _c-i\cos \frac \gamma 2\left| 2\right\rangle
_2\left| 0\right\rangle _c\stackrel{\text{(viii)}}{\rightarrow }\left( -\sin 
\frac \gamma 2\left| 0\right\rangle _2+\cos \frac \gamma 2\left|
1\right\rangle _2\right) \left| 1\right\rangle _c.
\end{eqnarray*}
\begin{equation}
\end{equation}
{\it \ }

Steps (ix)-(xi): Repeat the operations of Steps (iii)-(v) but set the
cavity-SQUID off-resonant interaction time as $\tau _6=$ $\beta \widetilde{%
\Delta }/\widetilde{g}^2$, leading to the transformation (18).

Step (xii): Adjust the level structure of SQUID $2$ to obtain off-resonant
interaction between the cavity mode and the $\left| 0\right\rangle
\leftrightarrow \left| 2\right\rangle $ transition of SQUID $2$ [Fig. 5(g)].
After an interaction time $\tau _7=\frac{\alpha \Delta }{\widehat{g}^2},$
the state $\left| 0\right\rangle _2\left| 1\right\rangle _c$ goes to $%
e^{i\alpha }\left| 0\right\rangle _2\left| 1\right\rangle _c$ while the
state $\left| 1\right\rangle _2\left| 1\right\rangle _c$ remains unchanged.

Step (xiii): Adjust the level structure of SQUID $2$ to achieve off-resonant
interaction between the cavity mode and the $\left| 1\right\rangle
\leftrightarrow \left| 2\right\rangle $ transition of SQUID $2$ [Fig. 5(h)].
After an interaction time $\tau _8=\frac{\alpha \Delta }{\widehat{g}^2},$
the state $\left| 1\right\rangle _2\left| 1\right\rangle _c$ changes to $%
e^{i\alpha }\left| 1\right\rangle _2\left| 1\right\rangle _c$ while nothing
happens to the state $\left| 0\right\rangle _2\left| 1\right\rangle _c.$

It can be seen that after the operations of Step (xii) and Step (xiii), the
transformation (19) is obtained as follows:

\begin{eqnarray}
&&\left| 0\right\rangle _2\left| 1\right\rangle _c\stackrel{\text{(xii)}}{%
\rightarrow }e^{i\alpha }\left| 0\right\rangle _2\left| 1\right\rangle _c%
\stackrel{\text{(xiii)}}{\rightarrow }e^{i\alpha }\left| 0\right\rangle
_2\left| 1\right\rangle _c,  \nonumber \\
&&\left| 1\right\rangle _2\left| 1\right\rangle _c\stackrel{\text{(xii)}}{%
\rightarrow }\left| 1\right\rangle _2\left| 1\right\rangle _c\stackrel{\text{%
(xiii)}}{\rightarrow }e^{i\alpha }\left| 1\right\rangle _2\left|
1\right\rangle _c.
\end{eqnarray}

Step (x-iv) Bring the $\left| 0\right\rangle \leftrightarrow \left|
2\right\rangle $ transition of SQUID $1$ to resonance with the cavity mode
for an interaction time $\tau _9=3\pi /\left( 2g_1\right) $ [Fig. 5(b)],
resulting in $\left| 0\right\rangle _1\left| 1\right\rangle _c\rightarrow
i\left| 2\right\rangle _1\left| 0\right\rangle _c.$

Step (x-v): Apply a $\pi $ microwave pulse with $\phi =\pi /2$ to SQUID $1$
[Fig. 5(a)]. The pulse is resonant with the $\left| 1\right\rangle
\leftrightarrow \left| 2\right\rangle $ transition of SQUID $1$. After the
pulse, the transformation $\left| 2\right\rangle \rightarrow \left|
1\right\rangle $ of SQUID $1$ is obtained.

One can see that the operations of Step (x-iv) and Step (x-v) lead to the
following transformation

\begin{equation}
\left| 0\right\rangle _1\left| 1\right\rangle _c\rightarrow i\left|
2\right\rangle _1\left| 0\right\rangle _c\rightarrow i\left| 1\right\rangle
_1\left| 0\right\rangle _c,
\end{equation}
which is actually the transformation (20) up to a phase factor $i$.

In above, we have explicitly shown how to realize the transformations
(15)-(20), i.e., performing a general $U$ on SQUID $2$ (the target) when
SQUID $1$ (the control) is initially in the state $\left| 1\right\rangle .$
On the other hand, it is noted that the following states of the system 
\begin{equation}
\;\left| 0\right\rangle _1\left| 0\right\rangle _2\left| 0\right\rangle
_c,\left| 0\right\rangle _1\left| 1\right\rangle _2\left| 0\right\rangle _c
\end{equation}
remain unchanged during the entire operation. This is because: (a) During
the operation of Step (i), the state $\left| 0\right\rangle $ of SQUID $1$
was not affected by the applied microwave pulse, since the $\left|
0\right\rangle \leftrightarrow \left| 2\right\rangle $ transition and the $%
\left| 0\right\rangle \leftrightarrow \left| 1\right\rangle $ transition of
SQUID $1$ are decoupled from the pulse; and (b) No photon was emitted to the
cavity during the operation of Step (ii), when SQUID $1$ is initially in the
state $\left| 0\right\rangle $. Hence, it can be concluded that the
transformation (14), i.e, the two-qubit controlled-$U$ gate (13) was
implemented with two SQUIDs after the above manipulation.

Before closing this section, several issues need to be addressed. The
irrelevant SQUIDs in each step of the operation need to be decoupled from
the cavity/pulse during the cavity/pulse-SQUID interaction. The cavity mode
needs to be not excited during the SQUID-pulse resonant interaction. In
addition, for each SQUID, the coupling of the levels $\left| 0\right\rangle $%
, $\left| 1\right\rangle ,$ and $\left| 2\right\rangle $ with the other
levels should be negligible. Note that for a SQUID, the level spacings can
readily be changed by varying the external flux $\Phi _x$ or the critical
current $I_c$ (e.g., for variable barrier rf SQUIDs) [33]. Therefore, these
conditions can in principle be satisfied by adjusting the level spacings of
the SQUIDs.

Imperfect decoupling between the irrelevant SQUIDs and the cavity during the
operations using off-resonant interaction could, in principle, result in
gate errors. Note that the population of the level $\left| j\right\rangle $
of any irrelevant SQUID initially in the state $\left| i\right\rangle $,
induced due to the coupling between the $\left| i\right\rangle
\leftrightarrow \left| j\right\rangle $ transition and the cavity mode, is
on the order of $p_j\simeq g_{ij}^2/(g_{ij}^2+\Delta _{ij}^2),$ where $%
i,j\in \{0,1,2\}$ and $i\neq j,$ $g_{ij}$ is the coupling constant of the
cavity mode with the $\left| i\right\rangle \leftrightarrow \left|
j\right\rangle $ transition of the SQUID, and $\Delta _{ij}=\omega
_{ij}-\omega _c$ is the detuning of the $\left| i\right\rangle
\leftrightarrow \left| j\right\rangle $ transition from the cavity mode.
Therefore, we remark that the coupling between the irrelevant SQUIDs and the
cavity can be made negligible as long as the condition $\Delta _{ij}>>g_{ij}$
and $\Delta _{ij}/g_{ij}^2>>\Delta /\widehat{g}^2,$ $\widetilde{\Delta }/%
\widetilde{g}^2$ can be satisfied. Here, $\widehat{g}$ and $\widetilde{g}$
are the off-resonant coupling constants described above, i.e., the coupling
constants of the cavity mode with the transition between the corresponding
two levels of the SQUID which is involved during the operation using
off-resonant interaction. The required decoupling condition can be obtained
by adjusting the level spacings of the irrelevant SQUIDs before the
off-resonant operations, so that the transition frequency between any two
levels of the irrelevant SQUIDs is highly detuned from the cavity-mode
frequency. It can be achieved with the available experiment technique
because the level spacings of a SQUID can be adjusted rapidly in experiment (%
$\sim $1ns). Since a more quantitative answer to the question of ``how well
decoupled the irrelevant SQUIDs need to be from the cavity'' requires a very
lengthy and complex analysis, we will not give a detailed discussion.

\begin{center}
{\bf IV. }${\bf N}$-{\bf QUBIT CONTROLLED-}$U${\bf \ GATES}
\end{center}

For $n$ qubits, there are a total number of $2^n$ computational basis states
from $\left| 00...0\right\rangle $ to $\left| 11...1\right\rangle $, which
form a set of complete orthogonal bases in a $2^n$-dimensional Hilbert space
of the $n$ qubits. As discussed in the introduction, an $n$-qubit controlled-%
$U$ gate performs an arbitrary unitary transformation $U$ on the target
qubit only when the $n-1$ control qubits are all in the state $\left|
1\right\rangle $, i.e., 
\begin{eqnarray}
\left| 1\right\rangle ^{\otimes \left( n-1\right) }\left| 0\right\rangle
&\rightarrow &\left| 1\right\rangle ^{\otimes \left( n-1\right) }U\left|
0\right\rangle ,  \nonumber \\
\left| 1\right\rangle ^{\otimes \left( n-1\right) }\left| 1\right\rangle
&\rightarrow &\left| 1\right\rangle ^{\otimes \left( n-1\right) }U\left|
1\right\rangle ,
\end{eqnarray}
while nothing happens to all other $2^n-2$ computational basis states. In
Eq. (27), the first $n-1$ qubits represent control qubits while the last
qubit acts as a target. In the following, we will discuss how this gate can
be achieved with $n$ SQUIDs coupled to a cavity.

The $n$ SQUIDs are labeled by $1,2,...,$ and $n.$ The first $n-1$ SQUIDs ($%
1,2,...,n-1$) represent control qubits while SQUID $n$ is the target qubit.
Suppose that all SQUIDs ($1,2,...,n$) are initially decoupled from the
cavity (which is in the vacuum state). We find that the $n$-qubit controlled-%
$U$ gate described above can be obtained through the following sequence of
operations (from right to left) 
\begin{equation}
U_1^{+}\otimes \left( \prod_{l=n-1}^1U_{lc}^{+}\right) \otimes U_{nc}\otimes
\left( \prod_{l=1}^{n-1}U_{lc}\right) \otimes U_1,
\end{equation}
where $\prod_{l=1}^{n-1}U_{lc}\equiv U_{(n-1)c}\cdot \cdot \cdot
U_{2c}U_{1c};$ $U_1$ denotes the operation on SQUID $1$ represented by
matrix 
\begin{equation}
U_1=\left( 
\begin{array}{cc}
0 & 1 \\ 
-1 & 0
\end{array}
\right)
\end{equation}
in the basis states $\left| 1\right\rangle _1=\left( 0,1\right) ^T$ and$%
\;\left| 2\right\rangle _1=\left( 1,0\right) ^T;$ $U_{lc}$ is a joint
operator on the SQUID $l$ and the cavity mode ($l=1,2,...,n-1$), represented
by the matrix 
\begin{equation}
\,U_{lc}=\left( 
\begin{array}{cc}
0 & -i \\ 
-i & 0
\end{array}
\right)
\end{equation}
in the basis states $\left| 0\right\rangle _l\left| 1\right\rangle _c=\left(
0,1\right) ^T$ and$\;\left| 2\right\rangle _l\left| 0\right\rangle _c=\left(
1,0\right) ^T;$ and $U_{nc}$ is a joint operator on SQUID $n$ and the cavity
mode, given by 
\begin{eqnarray}
U_{nc} &=&U\otimes I  \nonumber \\
&=&e^{i\alpha }R_z\left( \beta \right) R_y\left( \gamma \right) R_z\left(
\delta \right) \otimes I
\end{eqnarray}
which performs an arbitrary unitary transformation $U$ on the SQUID $n$
while nothing to the cavity state.

From the description in the previous section, it can be seen that:

(i) $U_1$ ($U_1^{+}$) can be realized by a $\pi $ microwave pulse ($\Omega
\tau _{\mu w}=\pi ,$ where $\tau _{\mu w}$ is the pulse duration) with $\phi
=-\pi /2$ ($\pi /2$) and $\omega _{\mu w}=\omega _{21}$ to SQUID $1;$

(ii) $U_{lc}$ corresponds to the operation of bringing the $\left|
0\right\rangle \leftrightarrow \left| 2\right\rangle $ transition of SQUID $%
l $ ($l=1,2...,n-1$) to resonance with the cavity mode for an interaction
time $\tau _l=\pi /\left( 2g_l\right) $;

(iii) $U_{nc}$ can be realized via the operations of Steps (iii)-(xiii)
described in the previous section. For the present case, the qubit involved
in the operations is SQUID $n$ instead of SQUID $2.$

(ii) $U_{lc}^{+}$ corresponds to the operation of bringing the $\left|
0\right\rangle \leftrightarrow \left| 2\right\rangle $ transition of SQUID $%
l $ ($l=1,2...,n-1$) to resonance with the cavity mode for an interaction
time $\tau _l=3\pi /\left( 2g_l\right) .$

To further understand Eq. (28), let us give some explanation on system
evolutions after the operations described above.

First, the state $\left| 1\right\rangle $ of SQUID $1$ is changed to the
ground state $\left| 0\right\rangle $ and a single photon is emitted to the
cavity mode after the joint operation $U_{1c}U_1$.

Second, after the joint operation $\prod_{l=2}^{n-1}U_{lc},$ the photon is
left in the cavity mode when SQUIDs ($2,3,...,n-1$) are initially in the
computational basis state $\left| 11...1\right\rangle ;$ however, it is
absorbed by SQUIDs ($2,3,...,n-1$) initially in all other computational
basis states.

Third, after the operation $U_{nc},$ an arbitrary unitary transformation $U$
is performed on SQUID $n$ with the assistance of the cavity photon; in
contrast, nothing happens to the SQUID $n$ if no photon, after the previous
operation $\prod_{l=2}^{n-1}U_{lc},$ is left in the cavity mode.

Fourth, after the joint operation $\prod_{l=2}^{n-1}U_{lc}^{+},$ the photon,
originally absorbed by SQUIDs ($2,3,...,n-1$), is emitted back to the cavity
mode.

Last, after the joint operation $U_1^{+}U_{1c}^{+},$ the cavity mode returns
to its original vacuum state $\left| 0\right\rangle _c$ and the state $%
\left| 0\right\rangle $ of SQUID $1$ is changed back to the initial state $%
\left| 1\right\rangle .$

It should be mentioned that when SQUID $1$ is initially in the state $\left|
0\right\rangle $, nothing happens to the whole system during the entire
operation, due to the reason discussed in the previous section.

To see it more clearly, let us consider the case of three qubits, i.e.,
realizing a three-qubit controlled-$U$ gate with three SQUIDs ($1,2,3$) in
cavity QED. For three qubits, there are a total number of eight ($2^3$)
computational basis states, denoted by $\left| 000\right\rangle ,\left|
001\right\rangle ,...,\left| 111\right\rangle ,$ respectively. The first
four basis states $\left| 000\right\rangle ,\left| 001\right\rangle ,\left|
010\right\rangle ,$ and $\left| 011\right\rangle $ of the three SQUIDs
remain unchanged during the operations described above. However, if the
three SQUIDs are initially in the other four basis states, the states of the
whole system after each unitary transformation of Eq. (28) ($n=3$) are: 
\[
\begin{array}{c}
\left| 100\right\rangle \left| 0\right\rangle _c \\ 
\left| 101\right\rangle \left| 0\right\rangle _c \\ 
\left| 110\right\rangle \left| 0\right\rangle _c \\ 
\left| 111\right\rangle \left| 0\right\rangle _c
\end{array}
\stackrel{U_1}{\longrightarrow } 
\begin{array}{c}
\left| 200\right\rangle \left| 0\right\rangle _c \\ 
\left| 201\right\rangle \left| 0\right\rangle _c \\ 
\left| 210\right\rangle \left| 0\right\rangle _c \\ 
\left| 211\right\rangle \left| 0\right\rangle _c
\end{array}
\stackrel{U_{1c}}{\longrightarrow } 
\begin{array}{c}
-i\left| 000\right\rangle \left| 1\right\rangle _c \\ 
-i\left| 001\right\rangle \left| 1\right\rangle _c \\ 
-i\left| 010\right\rangle \left| 1\right\rangle _c \\ 
-i\left| 011\right\rangle \left| 1\right\rangle _c
\end{array}
\stackrel{U_{2c}}{\longrightarrow } 
\begin{array}{c}
-\left| 020\right\rangle \left| 0\right\rangle _c \\ 
-\left| 021\right\rangle \left| 0\right\rangle _c \\ 
-i\left| 010\right\rangle \left| 1\right\rangle _c \\ 
-i\left| 011\right\rangle \left| 1\right\rangle _c
\end{array}
\]
\begin{equation}
\stackrel{U_{3c}}{\longrightarrow } 
\begin{array}{c}
-\left| 020\right\rangle \left| 0\right\rangle _c \\ 
-\left| 021\right\rangle \left| 0\right\rangle _c \\ 
-i\left| 01\right\rangle \left( U\left| 0\right\rangle \right) \left|
1\right\rangle _c \\ 
-i\left| 01\right\rangle \left( U\left| 1\right\rangle \right) \left|
1\right\rangle _c
\end{array}
\stackrel{U_{2c}^{+}}{\longrightarrow } 
\begin{array}{c}
-i\left| 000\right\rangle \left| 1\right\rangle _c \\ 
-i\left| 001\right\rangle \left| 1\right\rangle _c \\ 
-i\left| 01\right\rangle \left( U\left| 0\right\rangle \right) \left|
1\right\rangle _c \\ 
-i\left| 01\right\rangle \left( U\left| 1\right\rangle \right) \left|
1\right\rangle _c
\end{array}
\stackrel{U_{1c}^{+}}{\longrightarrow } 
\begin{array}{c}
\left| 200\right\rangle \left| 0\right\rangle _c \\ 
\left| 201\right\rangle \left| 0\right\rangle _c \\ 
\left| 21\right\rangle \left( U\left| 0\right\rangle \right) \left|
0\right\rangle _c \\ 
\left| 21\right\rangle \left( U\left| 1\right\rangle \right) \left|
0\right\rangle _c
\end{array}
\stackrel{U_1^{+}}{\longrightarrow } 
\begin{array}{c}
\left| 100\right\rangle \left| 0\right\rangle _c \\ 
\left| 101\right\rangle \left| 0\right\rangle _c \\ 
\left| 11\right\rangle \left( U\left| 0\right\rangle \right) \left|
0\right\rangle _c \\ 
\left| 11\right\rangle \left( U\left| 1\right\rangle \right) \left|
0\right\rangle _c
\end{array}
,
\end{equation}
where $\left| ijk\right\rangle $ is abbreviation of the state $\left|
i\right\rangle _1\left| j\right\rangle _2\left| k\right\rangle _3$ of SQUIDs
($1,2,3$) with $i,j,k\in \{0,1,2\}$. From Eq. (32), it can be seen that a
three-qubit controlled-$U$ gate is achieved with three SQUIDs, where the
third SQUID is the target qubit, after the transformations described in Eq.
(28).

On a final note, we point out that a single-mode cavity is not necessary
since for a multi-mode cavity one can in principle choose one mode to
interact with the SQUIDs while have all other modes well decoupled from the
three lowest levels of the SQUIDs. In addition, the method presented here is
applicable to a 1D, 2D, or 3D microwave resonator/cavity as long as the
conditions described above are satisfied.

\begin{center}
{\bf V. COMPARISON WITH CONVENTIONAL GATE CONSTRUCTION}
\end{center}

The universality of quantum computation implies that it is possible to
generate arbitrary $n$-qubit gates by using sequences of one-qubit and
two-qubit gates only [35-37]. Barenco {\it et. al. }have first developed
methods to design networks for $n$-qubit controlled gates [20]. They have
shown that it requires $2^{n-1}-1$ two-qubit controlled-$V$ and controlled-$%
V^{+}$ gates and $2^{n-1}-2$ two-qubit controlled-NOT gates to accomplish an 
$n$-qubit controlled-$U$ gate described above ($n\geq 3$). Here, $V$
satisfies $V^{2^{n-2}}=U.$ Namely, at least $2^n-3$ steps of operations are
required, assuming that realizing any two-qubit controlled gate requires
one-step operation only.

Since the work of Barenco et. al., much attention has been paid to optimal
implementation of quantum gates. Recently, M\"ott\"onen {\it et. al.} have
considered a generic elementary gate sequence for realizing a general
multiqubit gate [21]. More recently, Bergholm {\it et. al.} have presented
quantum circuits with uniformly controlled one-qubit gates [22]. According
to their results, $2^{n-1}-1$ two-qubit controlled-NOT gates, $2^{n-1}$
one-qubit gates, and a single diagonal $n$-qubit gate are needed to
construct the above $n$-qubit controlled-$U$ gate. Hence, at least $2^n$
steps of operations are required, provided that any two-qubit gate,
one-qubit gate, or an $n$-qubit diagonal gate can be implemented using
one-step operation only.

It is interesting to note that when compared with the use of the
conventional gate-decomposition protocols [20,22], our method significantly
reduces the number of operations needed to implement the above $n$-qubit
controlled-$U$ gate. As shown above, our method only needs $2n+11$ steps of
operations. The advantage of this method starts at $n=5$ and becomes more
dramatic as $n$ increases. It should be mentioned that the comparison
presented here is a conservative case because it was assumed above that a
two-qubit controlled-$V/$ $V^{+}/$NOT or a $n$-qubit diagonal gate can be
realized with one-step operation only.

\begin{center}
{\bf VI. DISCUSSION}
\end{center}

In this section, we give a brief discussion on relevant experimental issues
for the implementation of $n$-qubit controlled-$U$ gates. Without loss of
generality, let us consider $n$ identical SQUIDs ($1,2,...,n$) at locations
where the ${\bf B}_c$ fields are the same (e.g., antinodes of the cavity
field). Thus, we have $g_l=g$ ($l=1,2,...,n$). The total operation time is
given by 
\begin{equation}
\tau =\left[ 2n+\gamma /\left( 2\pi \right) \right] \tau _c^{(1)}+2\alpha
\tau _c^{\left( 2\right) }+(\beta +\delta )\tau _c^{\left( 3\right)
}+(2n+9)\tau _a+4\tau _{\mu w},
\end{equation}
where $\tau _c^{(1)}=\pi /g,\tau _c^{(2)}=\Delta /\widehat{g}^2,$ $\tau
_c^{\left( 3\right) }=\widetilde{\Delta }/\widetilde{g}^2,$ and $\tau _a$ is
the typical time required for adjusting the level spacings of a single
SQUID. For the method to work, $\tau $ should be much shorter than the
energy relaxation time $\gamma _2^{-1}$ of the level $\left| 2\right\rangle
, $ and the lifetime of the cavity mode $\kappa ^{-1}=Q/2\pi \nu _c,$ where $%
Q$ is the (loaded)\ quality factor of the cavity. In addition, direct
coupling between SQUIDs needs to be negligible since this interaction is not
intended.

These requirements can in principle be realized, since one can: (i) reduce $%
\tau _c^{(1)}$ by increasing the coupling constant $g$, (ii) shorten $\tau
_a $ by rapid adjustment of the level spacings of the SQUIDs, (iii) increase 
$\kappa ^{-1}$ by employing a high-$Q$ cavity so that the cavity dissipation
is negligible during the operation, and (iv) design SQUIDs and control
(readout) circuitry so that the energy relaxation time $\gamma _2^{-1}$ of
the level $\left| 2\right\rangle $ is sufficiently long. In addition, it is
noted that direct interaction between SQUIDs can be made negligible as long
as the following condition is satisfied 
\begin{equation}
M_{ss}\ll M_{sc},
\end{equation}
where $M_{ss}$ is the mutual inductance between two adjacent SQUIDs and $%
M_{sc}$ is the mutual inductance between each SQUID and the cavity.

For the sake of definitiveness, let us consider the experimental feasibility
of realizing a five-qubit controlled-$U$ gate using SQUIDs with the
parameters listed in Table 1. A five-qubit controlled-$U$ gate performs the
following transformation 
\begin{eqnarray}
\left| 1111\right\rangle \left| 0\right\rangle  &\rightarrow &\left|
1111\right\rangle U(\alpha ,\beta ,\gamma ,\delta )\left| 0\right\rangle , 
\nonumber \\
\left| 1111\right\rangle \left| 1\right\rangle  &\rightarrow &\left|
1111\right\rangle U(\alpha ,\beta ,\gamma ,\delta )\left| 1\right\rangle ,
\end{eqnarray}
when the four control qubits are in the state $\left| 1111\right\rangle $
while nothing otherwise. Here, $\gamma /2\in [0,2\pi ],$ and $\alpha ,\beta
/2,\delta /2\in [-\pi ,\pi ]$ (The positive or negative value for $\alpha
,\beta ,$and $\delta $ can be achieved by setting blue or red detuning).
Note that SQUIDs with the parameters in Table 1 are readily available at the
present time [2,3,38] and have the desired three-level structure as depicted
in Fig. 2. For a superconducting one dimensional standing-wave CPW (coplanar
waveguide) cavity with the parameters listed in Table 1 and SQUIDs placed
along the cavity axis (Fig. 6), one has $M_{sc}\sim 10^2$ pH. When each
SQUID is located at one of the antinodes of the cavity mode (Fig. 6), a
simple calculation gives $g\sim 5.8\times 10^9$ s$^{-1},$ resulting in $\tau
_c^{\left( 1\right) }\sim 0.5$ ns. On the other hand, as a rough estimate,
we assume $\widehat{g}\sim \widetilde{g}\sim 0.5g,$ $\Delta \sim 10\widehat{g%
}$, and $\widetilde{\Delta }\sim 10\widetilde{g},$ which can be readily
achieved by adjusting the level spacings. As a result, we have $\tau
_c^{\left( 2\right) }\sim \tau _c^{\left( 3\right) }\sim 3.4$ ns. With the
choice of $\tau _{\mu w}\sim \tau _a\sim \tau _c^{\left( 1\right) }$, one
has $\tau \sim 81.1$ ns for $\gamma /2=2\pi $ and $\alpha ,\beta /2,\gamma
/2,\delta /2=\pm \pi $ (a case requiring the longest operation time), which
is much shorter than $\gamma _2^{-1}\sim 3.2$ $\mu $s and $\kappa ^{-1}\sim
0.8$ $\mu $s for a cavity with $Q\sim 6\times 10^4$. Note that
superconducting CPW resonators with a quality factor of $Q>10^6,$ patterned
into a thin superconducting film deposited on the surface of a silicon chip,
has been experimentally demonstrated [39] (also see Refs. [23,26,27]
regarding its application for loaded superconducting qubits or semiconductor
qubits).

For a cavity with $\nu _c=11.4$ GHz, the wavelength of the cavity mode is $%
\lambda \sim 10.5$ mm. For each SQUID being placed at an antinode of the $%
{\bf B}_c$ field (Fig. 6), one has $D\sim 5.25$ mm, where $D$ is the
distance between the two nearest SQUIDs. A simple numerical calculation
gives $M_{ss}\sim 0.1$ aH, which is much smaller than $M_{sc}.$ Hence, the
condition of negligible direct coupling between SQUIDs is very well
satisfied.

Our above analysis demonstrates that the realization of a five-qubit
controlled-$U$ gate is possible using SQUIDs and a cavity within the present
technology. In addition, we point out that a quantum controlled Hadamard
gate with a larger number of control qubits can in principle be obtained by
increasing the length of the cavity though the conditions of $\tau \ll
\gamma _2^{-1},$ $\kappa ^{-1}$ becomes increasingly difficult to satisfy.

\begin{center}
{\bf VII. }$N$-{\bf QUBIT CONTROLLED-}$U${\bf \ GATES WITH ATOMS}
\end{center}

In this section, we discuss how to extend the above method to realize an $n$%
-qubit controlled-$U$ gate with three-level atoms, by the use of one cavity.

Consider $n$ identical atoms ($1,2,...,n$) each having $\Lambda $-type level
configuration formed by two ground states and an excited state (Fig. 7). In
accordance with the previous section, we use $\left| 0\right\rangle $ and $%
\left| 1\right\rangle $ to represent the two ground states and $\left|
2\right\rangle $ to indicate the excited state. The dipole transition
between $\left| 0\right\rangle $ and $\left| 1\right\rangle $ is forbidden
due to the definite parity of the wave function. In the following, the two
logical states of a qubit are represented by the two ground states $\left|
0\right\rangle $ and $\left| 1\right\rangle $ of each atom. And, atoms ($%
1,2,...,n-1$) act as control qubits while atom $n$ is the target qubit.

We note that an $n$-qubit quantum controlled-$U$ gate described by Eq. (27)
can be realized with three-level atoms using the prescription of Eq. (28),
by performing the following operations:

(a). Apply a $\pi /2$ classical pulse with $\phi =-\pi /2$ to atom $1,$
resonant with the $\left| 1\right\rangle \leftrightarrow \left|
2\right\rangle $ transition of atom $1.$ This pulse leads to the
transformation $\left| 1\right\rangle \rightarrow $ $\left| 2\right\rangle $
of atom $1,$ which accomplishes the transformation $U_1$ of Eq. (28).

(b). Send atom $1$ through the cavity. The cavity mode is initially in the
vacuum state $\left| 0\right\rangle _c$ and resonant with the $\left|
0\right\rangle \rightarrow $ $\left| 2\right\rangle $ transition of atom $1$
[Fig. 7(a)]. Choose the atomic velocity appropriately so that the passage
time of atom $1$ through the cavity equals to $\pi /\left( 2g_1\right) $.
Thus, after atom $1$ exits the cavity, the state $\left| 0\right\rangle
_1\left| 0\right\rangle _c$ remain unchanged, while the state $\left|
2\right\rangle _1\left| 0\right\rangle _c$ changes to $-i\left|
0\right\rangle _1\left| 1\right\rangle _c.$ This operation implements the
transformation $U_{1c}$ in Eq. (28).

(c). Send atoms ($2,3,...,n-1$) through the cavity one after another in a
way that no more than one atoms stay in the cavity simultaneously. The
cavity mode is resonant with the $\left| 0\right\rangle \rightarrow $ $%
\left| 2\right\rangle $ transition of each atom [Fig. 7(a)]. Choose the
atomic velocity appropriately so that the duration of atom $l$ in the cavity
equals to $\pi /\left( 2g_l\right) $ ($l=2,3,...n-1$). As a result, there is
no change for the state $\left| 0\right\rangle _l\left| 0\right\rangle _c,$ $%
\left| 1\right\rangle _l\left| 0\right\rangle _c,$ and $\left|
1\right\rangle _l\left| 1\right\rangle _c;$ while the state $\left|
0\right\rangle _l\left| 1\right\rangle _c$ becomes $-i\left| 2\right\rangle
_l\left| 0\right\rangle _c$. This process completes the transformation $%
\prod_{l=2}^{n-1}U_{lc}$ of Eq. (28).

(d). The transformation $U_{nc}$ in Eq. (30) is performed on atom $n$ and
the cavity mode, which is realized as follows:

(d.1) First, apply a $\pi /2$ classical pulse with $\phi =-\pi /2$ to atom $%
n,$ resonant with the $\left| 1\right\rangle \leftrightarrow \left|
2\right\rangle $ transition of atom $n$ and resulting in the transformation $%
\left| 1\right\rangle \rightarrow $ $\left| 2\right\rangle .$ Second, adjust
the cavity frequency [40] to obtain an off-resonant interaction between the
cavity mode and the $\left| 0\right\rangle \leftrightarrow \left|
2\right\rangle $ transition of atom $n$ [Fig. 7(b)] and then send atom $n$
through the cavity. After an interaction time $\frac{\delta \widetilde{%
\Delta }}{\widetilde{g}_n^2},$ the state $\left| 0\right\rangle _n\left|
1\right\rangle _c$ goes to $e^{-i\delta }\left| 0\right\rangle _n\left|
1\right\rangle _c$ while the state $\left| 2\right\rangle _n\left|
1\right\rangle _c$ changes to $e^{i\delta }\left| 2\right\rangle _n\left|
1\right\rangle _c$. Last, apply a $\pi /2$ pulse (with $\phi =\pi /2$) to
atom $n,$ resonant with the $\left| 1\right\rangle \leftrightarrow \left|
2\right\rangle $ transition of atom $1$ and resulting in the transformation $%
\left| 2\right\rangle \rightarrow $ $\left| 1\right\rangle .$ After these
operations, a rotation $R_z\left( \delta \right) $ on the two states $\left|
0\right\rangle $ and $\left| 1\right\rangle $ of atom $n$ is obtained while
the cavity mode remains in one-photon state.

(d.2) First, adjust the cavity frequency so that the cavity mode is resonant
with the $\left| 1\right\rangle \leftrightarrow \left| 2\right\rangle $
transition of atom $n$ [Fig. 7(c)]. Then send atom $n$ through the cavity
for an interaction time $\pi /\left( 2g_n^{\prime }\right) $ so that the
state $\left| 0\right\rangle _n\left| 1\right\rangle _c$ does not change
while the state $\left| 1\right\rangle _n\left| 1\right\rangle _c$ goes to $%
-i\left| 2\right\rangle _n\left| 0\right\rangle _c.$ Second, adjust the
cavity frequency so that the cavity mode is resonant with $\left|
0\right\rangle \leftrightarrow \left| 2\right\rangle $ transition of atom $n$
[Fig. 7(a)]. Then send atom $n$ back through the cavity for an interaction
time $\gamma /\left( 2g_n\right) $, leading to the rotation $\left|
0\right\rangle _n\left| 1\right\rangle _c\rightarrow \cos \frac \gamma 2%
\left| 0\right\rangle _n\left| 1\right\rangle _c-i\sin \frac \gamma 2\left|
2\right\rangle _n\left| 0\right\rangle _c$ and $\left| 2\right\rangle
_n\left| 0\right\rangle _c\rightarrow -i\sin \frac \gamma 2\left|
0\right\rangle _n\left| 1\right\rangle _c+\cos \frac \gamma 2\left|
2\right\rangle _n\left| 0\right\rangle _c.$ Last, have the cavity mode
resonant with the $\left| 1\right\rangle \leftrightarrow \left|
2\right\rangle $ transition of atom $n$ via the adjustment of the cavity
frequency [Fig. 7(c)]; and then send atom $n$ through the cavity for an
interaction time $3\pi /\left( 2g_n\right) .$ As a result, the state $\left|
0\right\rangle _n\left| 1\right\rangle _c$ remains unchanged while the state 
$\left| 2\right\rangle _n\left| 0\right\rangle _c$ becomes $i\left|
1\right\rangle _n\left| 1\right\rangle _c.$ It can be seen that after these
operations, a rotation gate $R_y\left( \gamma \right) $ on the two states $%
\left| 0\right\rangle $ and $\left| 1\right\rangle $ of atom $n$ is realized
in the same manner as shown in Eq. (23).

(d.3) To achieve a rotation $R_z\left( \beta \right) $ on the two states $%
\left| 0\right\rangle $ and $\left| 1\right\rangle $ of atom $n,$ one just
needs to perform the same operations as described in (d.1) by simply setting
the time of atom $n$ crossing the cavity as $\frac{\beta \widetilde{\Delta }%
}{\widetilde{g}_n^2}.$

(d.4) A common phase $e^{i\alpha }$ for the two states $\left|
0\right\rangle $ and $\left| 1\right\rangle $ of atom $n$ can be obtained as
follows. First, adjust the cavity frequency to obtain an off-resonant
interaction between the cavity mode and the $\left| 0\right\rangle
\leftrightarrow $ $\left| 2\right\rangle $ transition of atom $n$ [Fig.
7(d)] and then send atom $n$ through the cavity for an interaction time $%
\frac{\alpha \Delta }{\widehat{g}_n^2}$, resulting in $\left| 0\right\rangle
_n\left| 1\right\rangle _c\rightarrow $ $e^{i\alpha }\left| 0\right\rangle
_n\left| 1\right\rangle _c$ while no change for the state $\left|
1\right\rangle _n\left| 1\right\rangle _c.$ Second, adjust the cavity
frequency to obtain an off-resonant interaction between the cavity mode and
the $\left| 1\right\rangle \leftrightarrow \left| 2\right\rangle $
transition of atom $n$ [Fig. 7(e)]. Then, send atom $n$ back through the
cavity for an interaction time $\frac{\alpha \Delta ^{\prime }}{\widehat{g}%
_n^{\prime 2}},$ which leads to $\left| 1\right\rangle _n\left|
1\right\rangle _c\rightarrow e^{i\alpha }\left| 1\right\rangle _n\left|
1\right\rangle _c.$

The operations described above have accomplished a general transformation $U$
on the target qubit (atom $n$) with the assistance of the cavity photon.
However, {\it it is noted that the original states of the cavity mode and
atoms (}$1,2,...n-1${\it ) have also changed after the above operations.}
Therefore, it is necessary to return the cavity mode and atoms ($1,2,...,n-1$%
) (control qubits) to their original states, which can be done through the
following operations:

(e) Send atoms ($2,3,...,n-1$) through the cavity one after another, without
more than one atoms staying in the cavity simultaneously. The cavity
frequency is adjusted to have the cavity mode resonant with the $\left|
0\right\rangle \leftrightarrow \left| 2\right\rangle $ transition of each
atom [Fig. 7(a)]. Choose the atomic velocity appropriately so that the
duration of atom $l$ in the cavity equals to $3\pi /\left( 2g_l\right) $ ($%
l=2,3,...n-1$). As a result, nothing happens to the states $\left|
0\right\rangle _l\left| 0\right\rangle _c$ and $\left| 1\right\rangle
_l\left| 0\right\rangle _c$ while the state $\left| 2\right\rangle _l\left|
0\right\rangle _c$ becomes $i\left| 0\right\rangle _l\left| 1\right\rangle
_c $ ($l=2,3,...n-1$). After this operation, atoms ($2,3,...n-1$) return to
their original states and the photon, originally absorbed by atoms ($%
2,3,...,n-1$), is emitted back to the cavity. This process realizes the
transformation $\prod_{l=n-1}^2U_{lc}^{+}$ of Eq. (28).

(f). Send atom $1$ back through the cavity. The cavity mode is resonant with
the $\left| 0\right\rangle \leftrightarrow \left| 2\right\rangle $
transition of atom $1$ [Fig. 7(a)]. Choose the atomic velocity appropriately
so that the passage time of atom $1$ through the cavity equals to $3\pi
/\left( 2g_1\right) .$ Thus, after atom $1$ exits the cavity, the state $%
\left| 0\right\rangle _1\left| 0\right\rangle _c$ remains unchanged but the
state $\left| 0\right\rangle _1\left| 1\right\rangle _c$ changes to $i\left|
2\right\rangle _1\left| 0\right\rangle _c.$ After this operation, the cavity
mode returns to the original vacuum state $\left| 0\right\rangle _c$ and the
transformation $U_{1c}^{+}$ of Eq. (28) is obtained.

(g). Finally, apply a $\pi /2$ classical pulse (with $\phi =\pi /2$) to atom 
$1.$ The pulse is resonant with the $\left| 1\right\rangle \leftrightarrow
\left| 2\right\rangle $ transition of atom $1.$ This pulse leads to $\left|
2\right\rangle \rightarrow $ $\left| 1\right\rangle ,$ completing the
transformation $U_1^{+}$ of Eq. (28).

The present scheme has the following advantages:

(i) No adjustment of the level spacings for each atom is required during the
operations;

(ii) Only one cavity is required;

(iii) No identical atom-cavity coupling constants are needed; and

(iv) The total number of basic operations is $2n+11,$ which is much less
than that required by the conventional gate-decomposition protocols [20,22]
when $n$ is a larger number ($n\geq 5$).

\begin{center}
{\bf VIII. CONCLUSION}
\end{center}

We have presented a method to realize a multiqubit quantum controlled-$U$\
gate with SQUIDs coupled to a microwave cavity. The method operates
essentially by creating a single photon through one of the control SQUIDs
and then exchanging the photon between the control SQUIDs and the cavity
mode before and after a unitary transformation $U$ is performed on the
target SQUID. The method has these advantages: (i) Since no tunneling
between the qubit levels $\left| 0\right\rangle $ and $\left| 1\right\rangle 
$ is required, decay from the level $\left| 1\right\rangle $ can be made
negligibly small during the operation, via prior adjustment of the barrier
of the double-well potential [33]; (ii) As neither measurement on
SQUIDs/photons nor auxiliary SQUID is needed, the operation is simplified
and hardware resources is saved; (iv) Because coupling constants of SQUIDs
with the cavity are not required to be identical, inevitable nonuniformity
in device parameters is tolerable and non-exact placement of SQUIDs is
allowed; (iv) The method can in principle be applied to obtain an $n$-qubit
controlled-$U$ gate with a large number $n$, and (v) More interestingly, the
gate operations are significantly simplified as the number of qubits
increases, when compared with the use of the conventional gate-decomposition
protocols. As shown above, the present method can be extended to implement a
multiqubit controlled-$U$ gate with atoms in cavity QED. Finally, it is
noted that the method is also applicable to the realization of a multiqubit
controlled-$U$ gate with quantum dots in cavity QED [41].

Before we conclude, it should be mentioned that the idea of realizing
multiqubit controlled phase gates with superconducting flux qubits or charge
qubits has been proposed previously [42,43]. Our present work, however,
deals with the realization of a multiqubit controlled-$U$ gate\ (a
multiqubit controlled-``{\it arbitrary transformation}''). Therefore, it is
much more general than the previous works [42,43]. To the best of our
knowledge, no one has yet demonstrated how to realize an $n$-qubit
controlled-$U$ gate within cavity QED. We believe that this work is of great
importance since it provides a simple protocol to realize a multiqubit
controlled-$U$ gate with SQUID qubits or atomic qubits within cavity QED.

\begin{center}
{\bf ACKNOWLEDGMENTS}
\end{center}

This work was partially supported by National Science Foundation ITR program
(DMR-0325551), and AFOSR (F49620-01-1-0439), funded under the Department of
Defense University Research Initiative on Nanotechnology (DURINT) Program
and by the NSA.

\begin{center}
{\large Figure and Table Captions\\}
\end{center}

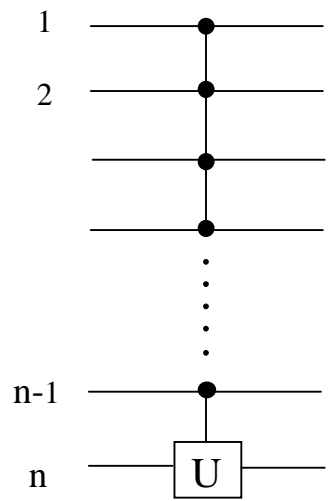
FIG. 1. Schematic circuit of an $n$-qubit controlled-$U$ gate. A unitary
transformation $U$ is performed on the target qubit (qubit $n$) when the $%
n-1 $ controls on the filled circles (qubits $1,2,...,$ and $n-1$) are all
in the state $\left| 1\right\rangle $.

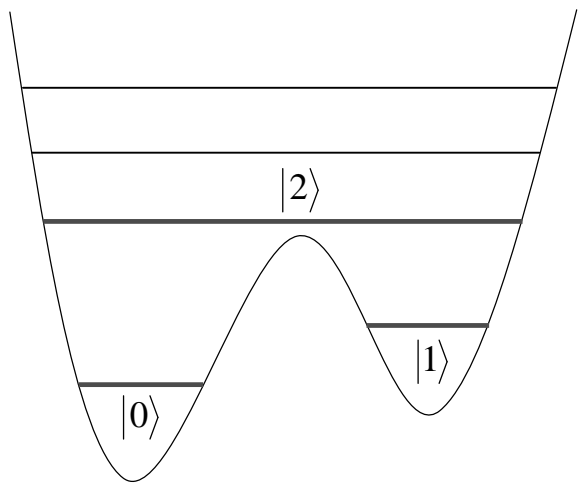
FIG. 2. Level diagram of a SQUID with the $\Lambda $-type three lowest
levels $\left| 0\right\rangle ,$ $\left| 1\right\rangle $ and $\left|
2\right\rangle $.

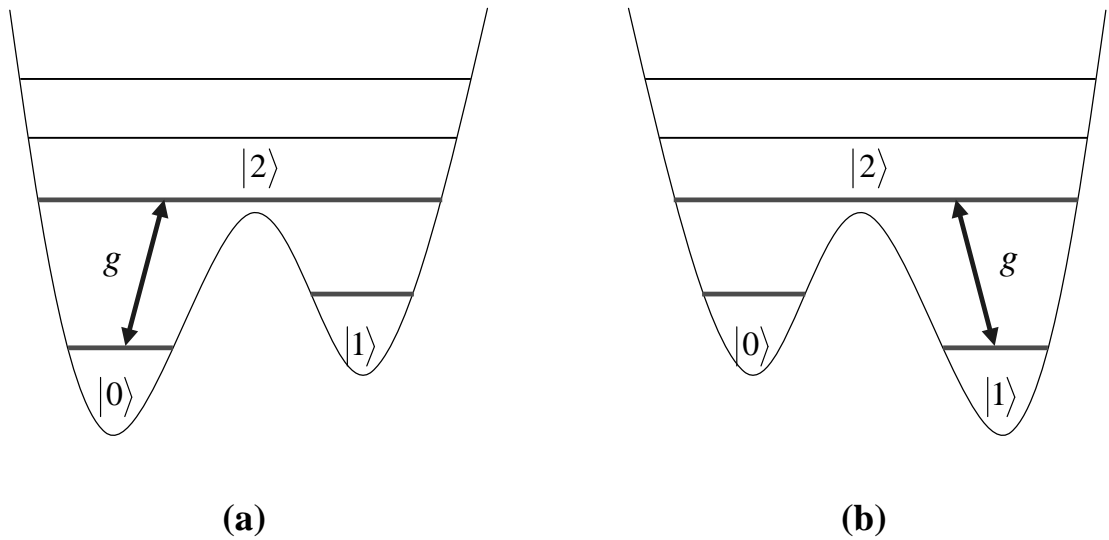
FIG. 3. (a) Resonant interaction of the cavity mode with the $\left|
0\right\rangle \leftrightarrow \left| 2\right\rangle $ transition of a
SQUID. (b) Resonant interaction of the cavity mode with the $\left|
1\right\rangle \leftrightarrow \left| 2\right\rangle $ transition of the
same SQUID. In (a) and (b), the coupling constant $g$ is the same. Figure
(b), where the ground level is $\left| 1\right\rangle $ and the first
excited level is $\left| 0\right\rangle ,$ is obtained by flipping Figure
(a). The potential and the level structure shown in (a) and (b) can be
obtained with external flux bias of $\left( 0.5-\delta \right) \Phi _0$ and $%
\left( 0.5+\delta \right) \Phi _0,$ respectively.

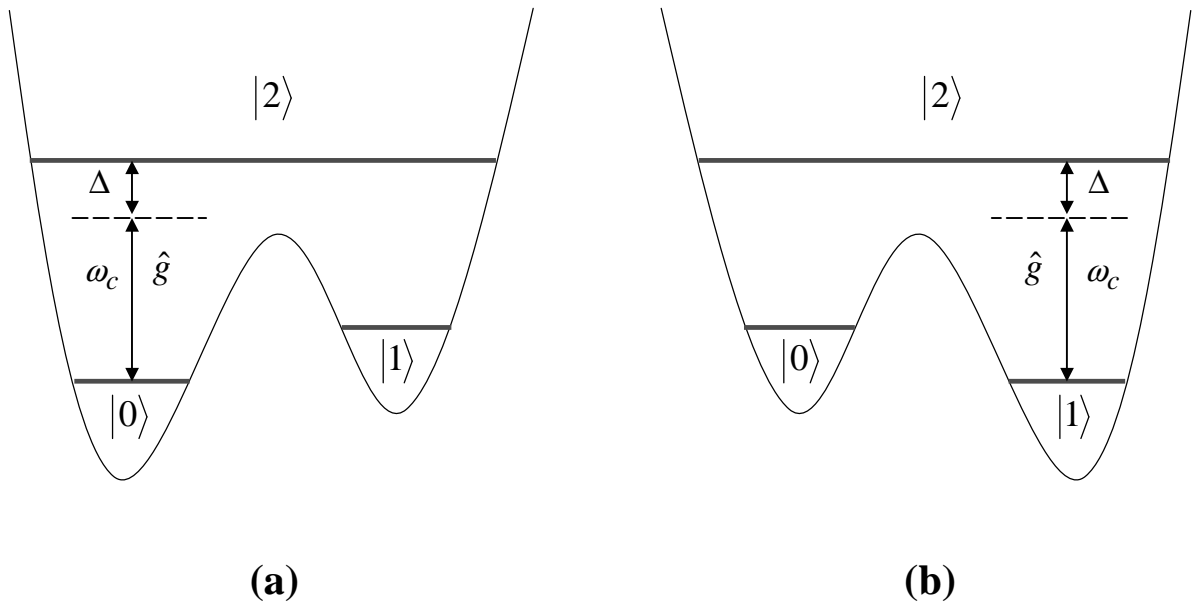
FIG. 4. (a) Off-resonant interaction between the cavity mode and the $\left|
0\right\rangle \leftrightarrow \left| 2\right\rangle $ transition of a
SQUID. (b) Off-resonant interaction between the cavity mode and the $\left|
1\right\rangle \leftrightarrow \left| 2\right\rangle $ transition of the
same SQUID. The coupling constant $\widehat{g}$ and the detuning $\Delta $
in (b) are the same as those in (a), which can be achieved with external
flux bias of $\left( 0.5-\delta \right) \Phi _0$ and $\left( 0.5+\delta
\right) \Phi _0$ for (a) and (b), respectively. In (a) $\Delta $ $=\omega
_{20}-\omega _c,$ while in (b) $\Delta $ $=\omega _{21}-\omega _c.$

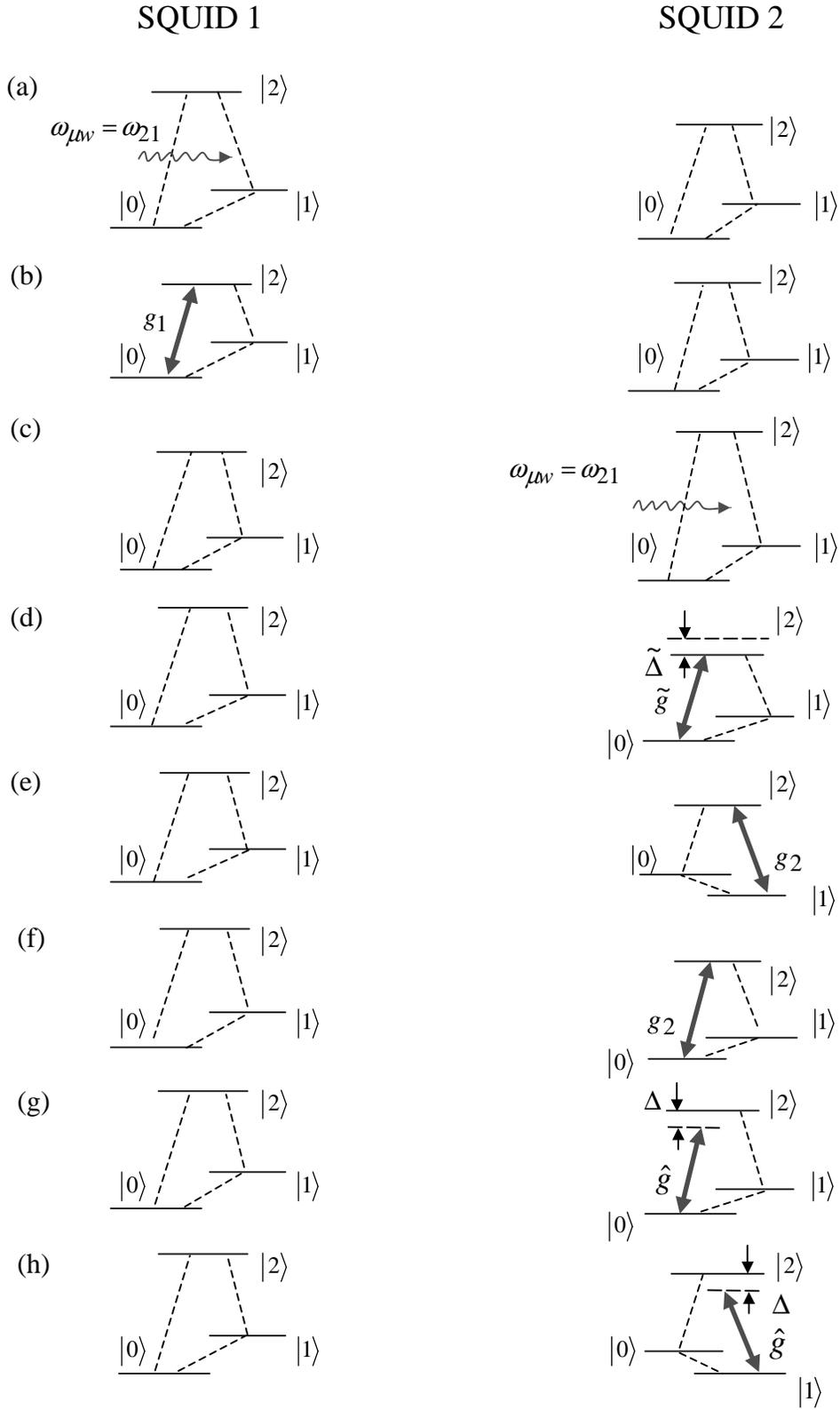
FIG. 5. Change of the level structure (reduced) of SQUIDs ($1,2$) during a
two-qubit controlled-$U$ gate performance. In (a), (b), (c), (d), (e), (f),
(g), and (h), figures from left to right represent the level structures for
SQUIDs $1$ and $2$, respectively; the non-identical level spacings of the
SQUIDs could be caused by nonuniform device parameters. In (a) and (c), the
level spacings for the two SQUIDs are set to be much different, such that
the irrelevant SQUID is decoupled from the applied pulse. The transition
between any two levels linked by a dashed line is decoupled from the cavity
mode. $\widetilde{g}$ and $\widehat{g}$ are the off-resonant coupling
constants between the cavity mode and the corresponding two-level transition
of SQUID $2.$ $g_1$ and $g_2$ are the SQUID-cavity resonant coupling
constants for SQUID 1 and SQUID 2, respectively. The detuning $\Delta =$ $%
\omega _{20}-\omega _c$ for (g) while $\omega _{21}-\omega _c$ for (h). In
addition, $\widetilde{\Delta }=$ $\omega _c-\omega _{20}.$

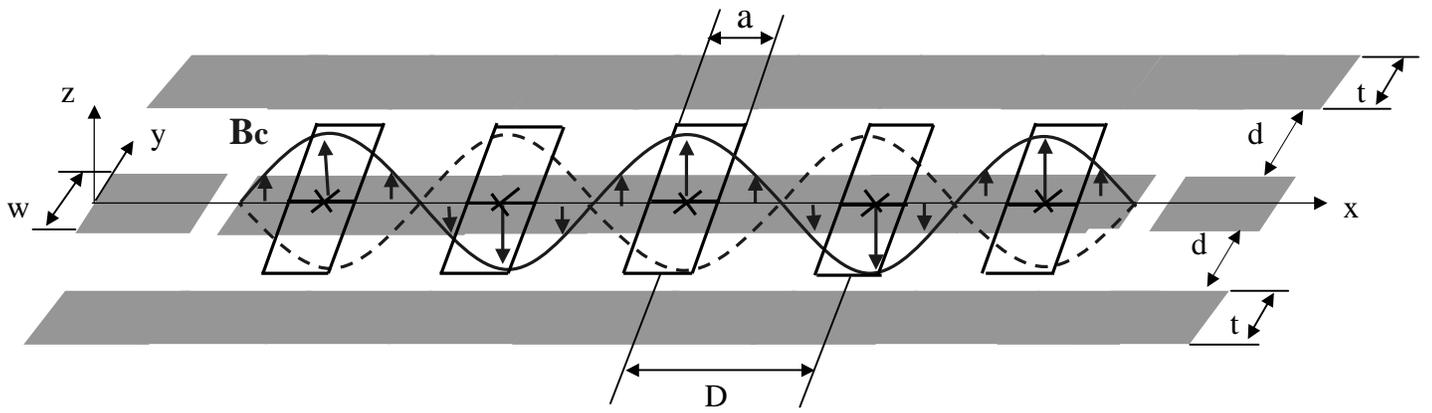
FIG. 6. Sketch of the setup for five SQUIDs ($1,2,3,4,5$) and a
standing-wave quasi-one dimensional CPW cavity (Not drawn to scale). Each
SQUID is placed in the plane of the resonator between the two lateral ground
planes (i.e., the $x$-$y$ plane) and at an antinode of the ${\bf B}_c$
field. The two curved lines represent the standing-wave ${\bf B}_c$ field,
which is in the $z$-direction.

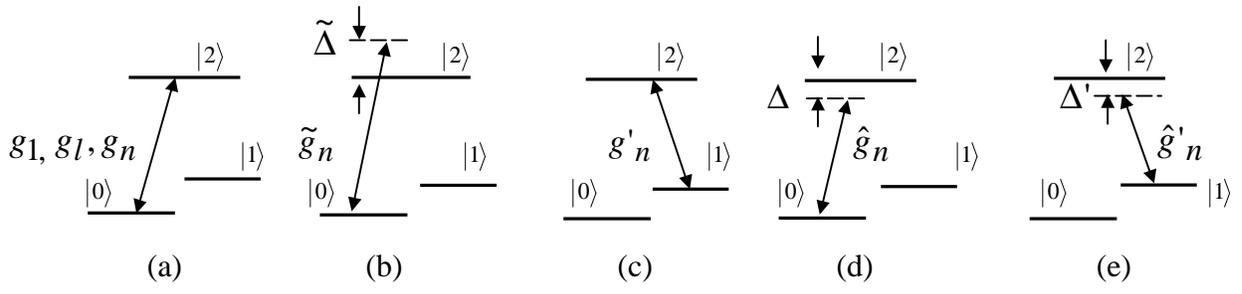
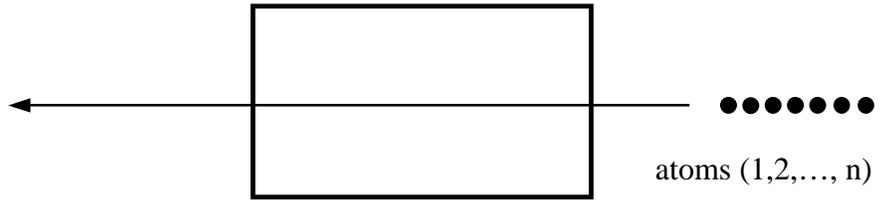
FIG. 7. Sketch of the setup for the realization of a controlled-$U$ gate
with three-level atoms and a cavity. (a) Resonant interaction of the cavity
mode with the $\left| 0\right\rangle \leftrightarrow \left| 2\right\rangle $
transition of atom $1,$ atom $l$ ($l=2,3,...,n-1$), or atom $n$. (c)
Resonant interaction of the cavity mode with the $\left| 1\right\rangle
\leftrightarrow \left| 2\right\rangle $ transition of atom $n$. (b), (d)
Off-resonant interaction of the cavity mode with the $\left| 0\right\rangle
\leftrightarrow \left| 2\right\rangle $ transition of atom $n$ for different
detuning setting. (e) Off-resonant interaction of the cavity mode with the $%
\left| 1\right\rangle \leftrightarrow \left| 2\right\rangle $ transition of
atom $n.$ $g_1$, $g_l$, and $g_n$ are the resonant coupling constants
between the cavity mode and the $\left| 0\right\rangle \leftrightarrow
\left| 2\right\rangle $ transition of atom $1$, atom $l$ ($l=2,3,...,n-1$),
and atom $n,$ respectively. $g_n^{\prime }$ is the resonant coupling
constant between the cavity mode and the $\left| 1\right\rangle
\leftrightarrow \left| 2\right\rangle $ transition of atom $n.$ $\widetilde{g%
}_n,\widehat{g}_n$ and $\widehat{g}_n^{\prime }$ are the off-resonant
coupling constants between the cavity mode and the corresponding two-level
transition of atom $n.$ $\Delta =\omega _{20}-\omega _c,$ $\Delta ^{\prime
}=\omega _{21}-\omega _c,$ and $\widetilde{\Delta }=\omega _c-\omega _{20}.$

TABLE 1. Parameters for a SQUID-cavity. $\beta _L$ is the SQUID's potential
shape parameter, $R$ is the SQUID's effective damping resistance, and $S$ is
the surface bounded by the loop of the SQUID with width $a$ and length $b$. $%
\gamma _2^{-1}$ ($\gamma _1^{-1}$) is the energy relaxation time of the
level $\left| 2\right\rangle $ ($\left| 1\right\rangle $). $\nu _{20}$ ($\nu
_{21}$) is the $\left| 0\right\rangle \leftrightarrow \left| 2\right\rangle $
($\left| 1\right\rangle \leftrightarrow \left| 2\right\rangle $) transition
frequency. $\phi _{ij}\equiv \left\langle i\right| \Phi \left|
j\right\rangle /\Phi _0$ is the magnetic dipole coupling matrix element
between levels $\left| i\right\rangle $ and $\left| j\right\rangle $ ($%
i=1,2; $ $j=0,1$). $l$ is the length of the quasi-one dimensional CPW
cavity, $\lambda $ is the wavelength of the cavity mode with frequency $\nu
_c,$ $d$ is the gap between the center conductor and the adjacent ground
plane, $w$ is the width of the center conductor, $t$ is the width of each
ground plane, $L_0$ is the inductance per unit length of the waveguide, and $%
\varepsilon _e $ is the effective relative dielectric constant.

\end{document}